# STUDY OF ISOSCALAR MESONS IN NUCLEAR MEDIUM

G.A. Sokol [1], E.M. Leikin [2]
1. Lebedev Phys.Institute RAS, e-mail: gsokol@venus.lpi.troitsk.ru
2. Nucl.Phys.Institute,Moscow State University,email:leikin@sinp.msu.ru

Abstract

Experimental studies of possible modifications of isoscalar-meson properties in nuclear medium are proposed.

Some papers [1,2,3,4] report a shift in energy distribution of decay particles in a nuclei relative that corresponding to free meson decay. The effect observed had been considered as possible influence of nuclear media on meson properties. This changing of meson properties is known to be associated with partial restoration of spontaneous breaking of chiral symmetry [5]. Following is discussion of an accelerator experiment for direct proof of meson mass changing in nuclear medium.

The following conditions have to be fulfilled for such experiment:

1. The specific types of mesons produced by an incident particle in nuclei have to be chosen. For an experiment, we select mesons with pronounced decay in two particles – photons or any other charged particles. This provides the most confident registration of such decay products during measurements.

2. The recoilless kinematics has to be selected. Namely, the velocity of meson produced has to be close to zero in order to guarantee particle decay inside the nuclei. In contrast to meson production in interaction of incident beam with free nucleon, in nuclei recoilless kinematics condition can be satisfied approximately only due to nucleon Fermi motion inside nuclei. In such a kinematics, decay particles are distributed uniformly in full angle, and spectrometer position might be chosen from the low background considerations.

3. Target material has to contain protons besides main nucleus. This provides direct measurements of meson mass shift in nuclei.

The mesons η, ω, ή and φ are possible candidates for their decay exploration in nuclei [6]. Table 1 illustrates these particle properties, while table 2 represent the reactions used for meson mass shift experiment. As it is seen, gamma, electron, proton as well as deutron beams may be used. It is worthwhile to pay attention on resonance $S_{11}$ (1535) mass shift in experiments on so called η-nucleus [7,8,9].

The reason for composite target ($CH_2$ or $C_3H_8$ films, for a example) is the following. Multi nucleon nuclei provides nuclear media for meson production and decay, thus making it possible meson mass shifting relative to that produced in reaction on free proton. Two maxima in total energy spread as it can be seen from fig.1 are responsible for different processes, and ΔE is expected shift to be measured. It is clear as well that the narrower appropriate resonance width the better resolution.

More detailed calculations are necessary and foreseen concerning general considerations have been made in this note

Authors thank for help in this work V.G. Kurakin and V.A.Baskov .

Table 1. Main characteristics and 2 particles branching ( Particle Data - 2008 )

| meson | mass (MeV) | width Γ(MeV) | decay | branching % | p (MeV/c) |
|---|---|---|---|---|---|
| η (547) | 547,85 ± 0,02 | 1,30 ± 0,07 (KeV) | γγ | 39,31 ± 0,20 | - |
| ω (782) | 782,85 ± 0,12 | 8,49 ± 8,49 | π+π- | 1,53 ± 0,12 | 366 |
| ή (958) | 957,66 ± 0,24 | 0,205 ± 0,015 | γγ | 2,10 ± 0,12 | 479 |
| φ (1020) | 1019,45 ± 0,18 | 4,26 ± 0,04 | K+K- | 49,20 ± 0,6 | 127 |

Table 2. Proton kinetic energies T(p) and gamma energies E(γ) for meson production at recoilless condition T(M) = 0 (GeV)

|  | η<br>0,547 | ω<br>0,782 | ή<br>0,958 | φ<br>1,020 |
|---|---|---|---|---|
| T(p)  p + p = M + p + p | 2,24 | 7,37 | - | - |
| T(p)  p + d = M + p + d | 1,04 | 1,73 | 2,43 | 2,74 |
| E(γ)  γ + p = M + p | 1,0 | 1,90 | - | - |
| E(γ)  γ + d = M + d | 0,63 | 0,95 | 1,20 | 1,42 |

Table 3. Threshold energies (GeV)

|  | η<br>0,547 | ω<br>0,782 | ή<br>0,958 | φ<br>1,020 |
|---|---|---|---|---|
| E(γ)  γ + p = M + p | 0,71 | 1,11 | 1,45 | 1,57 |
| T(p)  p + p = M + p + p | 1,25 | 1,89 | 2,40 | 2,59 |

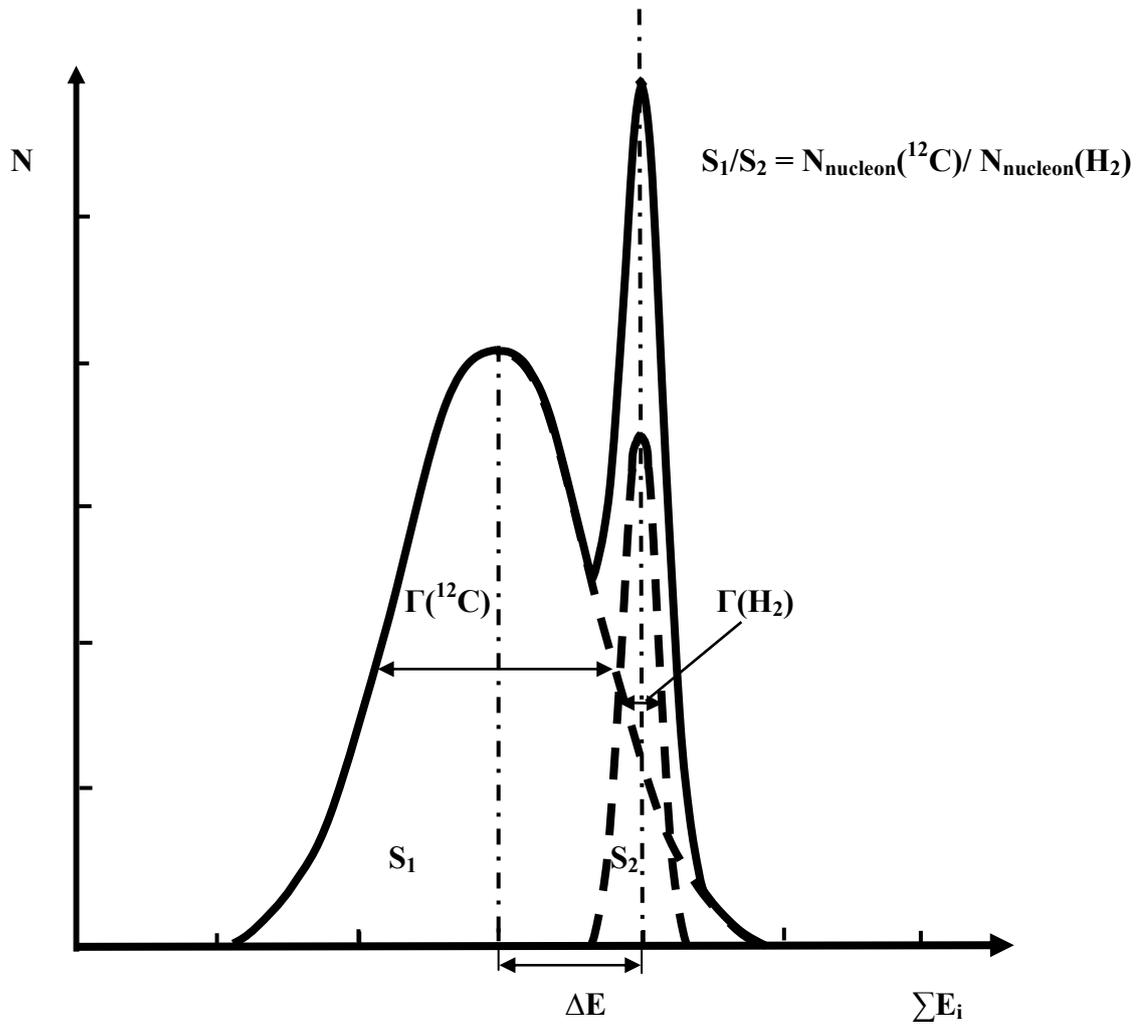

Fig.1. Expected distribution of sum energy particles $\sum E_i$ from decay M meson produced on composite target $CH_2$ is schematically displayed